\documentclass[12pt,english,american,aps,manuscript]{revtex4}
\usepackage[T1]{fontenc}
\usepackage[latin1]{inputenc}
\usepackage{amssymb}

\makeatletter

\usepackage{babel}
\makeatother
\begin{document}

\author{Jakob Lamey and Gustav M. Obermair}

\date{28 November 2005}

\address{{\footnotesize Department of Physics University of Regensburg D-93040
Regensburg, Germany }}

\email{jakob.lamey@physik.uni-regensburg.de}

\begin{abstract}
In what respect does terrestrial physics reflect the two unique features
of the global gravitational field: its infinite range and its equivalence
to spacetime curvature? We quote the evidence that true irreversibility,
i.e. the growth of the Boltzmann entropy of any finite system, is
the consequence of the global state of the gravitation dominated and
expanding universe. Moreover, as another example, we calculate the
effect of global expansion and of the gravitational potential observed
in our Local Group on space-time metric in terms of curvature. Surprisingly,
we find an energy density $T^{00}$ which is in numerical agreement
with the purely quantum theoretical result for the Casimir energy
density containing Planck's constant.
\end{abstract}

\title{Local effects of global gravitation}


\maketitle

\section{Introduction}

Exploring the dynamics of our solar system in the seventeenth century
and the discovery of its governing principle - the universal law of
gravitation - marks the beginning of modern science and has paved
the way towards enlightenment. Yet even 300 years after Newton and
90 years after Einstein's general relativity we seem still to be just
at the beginning of gravitational physics and still search for the
connection between our theories of the quantum dominated micro-world
and the gravitation dominated universe. At least in one case - that
of the cosmological constant - the astronomically observed value is
in horrendous discrepancy with predictions from micro-physics \cite{adleretalcosmolconstamj,carollcosmolconstpluscomm}.
If Newton himself would have tried to grasp as a total whole what
was observable in the universe at his time, perhaps today we would
still wait for the knowledge about gravity we learned from him. And
many new discoveries would not have been made.

For the present understanding of gravitation a basic discovery is
that of physical fields having a reality of their own. It freed us
from the perception of an absolute space devoid of any interaction
- a global inertial system. What Einstein wrote in 1953 is valid until
today \cite{jammerspace}: ''Another way to overcome the inertial
system than by field theory no one has found so far.'' In this paper
we take the existence and properties of the global gravitational field
for granted: its infinite range and the curvature induced by - or
rather equivalent to - it and look for local manifestations of these
properties. In Section II this is done in the context of statistical
physics along an evidence not often found in textbooks on statistical
mechanics. In Section III we discuss large scale observations and
their quantum physical relevance. The values of these large scale
observations are used in Section IV to express quantum physical orders
of magnitudes solely by the field theory of gravitation - Einstein's
general relativity.

\section{Gravitation and the second Law}

The impact of gravitation on the statistical behavior of local, e.g.
terrestrial macro-systems can be seen as follows: The \emph{statistical}
or Boltzmann-Gibbs entropy $\,\,\,\,-\left\langle \ln\rho\right\rangle $
is easily shown to be strictly constant in time both in classical
and in quantum statistical mechanics as long as the density $\rho$
and the average $\left\langle \ldots\right\rangle $ refer to a strictly
closed system. It is only the macroscopic or thermodynamic entropy
of systems prepared in a non-equilibrium state, i.e. their ability
to do macroscopic work, that does in fact exhibit an often rather
fast increase, even when they are practically isolated. This, however,
has nothing to do with fundamental irreversibility, i.e. with a loss
of order on the level of micro-states, but is due to reversible, but
ergodic and mixing dynamics and to the coarse graining resulting inevitably
from the macroscopic nature of all measurements.

Where, then, lies the cause of the loss of microscopic order, of true
time asymmetry and of the universal arrow of time that apparently
prevails even though interactions are time-symmetric? It lies in the
fact that with respect to gravitation no system is ever closed. The
bigger we make a system in an attempt to include its ''walls'',
the more it is dominated by gravitation, the more open it gets: we
live in an open universe. As already pointed out by Borel 80 years
ago \cite{ZehBor}, moving a mass of some ${\rm kg}$ by some ${\rm m}$
on Sirius 5 light-years away would completely change the micro-state
of $10^{23}$ molecules in a container on earth within a few minutes.
The farthest boundary about which empirical science may talk is the
horizon where the red-shift diverges; actio in the sense of Newton's
3rd law may go out towards it, but reactio does not come back from
beyond. Low entropy states of subsystems (hot coffee in a cool room,
living beings in cold space) are possible, because the one universe
in which we (are able to) live started in a state of extreme order,
this order slowly being sucked out towards infinity. Thus the growth
of disorder observed in any finite subsystem can proceed through time-symmetric
interactions, not because this is a law of nature but because it is
a property of the very special state of our universe. This insight
is certainly not completely new, but seems to find its way only very
slowly into statistical mechanics textbooks.

\section{Equivalence principle and large scale observations}

Does the present state of our universe produce a gravitational environment,
which has not been quantified in terms of field theory so far? Indeed,
two large scale astronomical observations can be connected via the
specific quality of the gravitational field, the equivalence principle.
The first is the observation of Hubble expansion being accelerated
globally, the second is the observation of the Local Group being accelerated
locally. If according to the equivalence principle a uniform acceleration
locally corresponds to a constant homogeneous gravitational field,
the accelerated Hubble expansion must have a gravitational counterpart.
This gravitational counterpart might be given by the gravitational
potential which is responsible for the acceleration of the Local Group.
The absolute value of such a potential $\phi$ in the solar system
can be estimated from astronomical observations \cite{lyndbellea88,tonryetalsbfII}
to be about $\phi\approx10^{-5}$ in geometric units. The scales are
such that $\phi$ is constant in space through the entire solar system
up to $10^{-11}$. The absolute value of $\phi$ lies well above the
value of the potential of the sun on earth and well above linear approximations
of general relativity \cite{misnerthornewheeler73}. 

The idea that this potential might get significance in quantum physics
has been pushed forward by Ahluwalia in a seminal paper \cite{ahluwgrcompl}.
Similar to the Aharonov-Bohm effect \cite{aharonovbohm} showing in
the case of electromagnetic interaction that it is quantum physics
which in contrast to classical physics is able to detect even a constant
potential itself, quantum physics should also be able to detect a
regionally constant potential in the case of gravitational interaction.
This argument is underlined in \cite{ahluwgrcompl} by the fact that
the gradient of $\phi$ is practically zero through the entire solar
system. In \cite{ahluwgrcompl} the role of $\phi$ in non-relativistic
quantum mechanics is discussed and its effect on neutrino oscillations
is estimated. In both cases a linear, weak field approximation of
general relativity is used implying a flat background. Here, we discuss
the physical significance of $\phi$ from a general relativistic viewpoint,
because its high absolute value makes a mathematical treatment in
a flat inertial background questionable.

\section{Gravitation and local curvature}

According to general relativity a gravitational field acts as curvature
of spacetime. If there is no significant spatial variation of the
field within the solar system, it is only temporal variation, which
remains in order to make such a potential accessible to differential
geometry in a spacetime approach. This temporal variation is given
by the Hubble expansion. Indeed, as is shown in \cite{bjp}, the present
order of magnitude of the Hubble constant and that of such a gravitational
potential fit together surprisingly well in fairly simple curvature
calculations. Here we do the calculations without any reference to
a specific cosmological model, especially without use of the deceleration
parameter $q_{0}$. 

The potential gains time dependence by the scale factor of the Hubble
expansion,\begin{equation}
\phi(t)=-\frac{r_{g}}{l(t)}\ \mbox{and}\ \dot{\phi}(t)=\frac{r_{g}}{l(t)}\frac{\dot{l}(t)}{l(t)}=-\phi(t)H(t),\label{eq:tdepscalepot}\end{equation}
where $r_{g}=\frac{GM}{c^{2}}$ denotes the gravitational radius and
$H(t)=\frac{\dot{l}(t)}{l(t)}$ the Hubble function. Then a purely
time-dependent spacetime metric can be constructed\[
g_{\mu\nu}={\rm diag}(-A(t),B(t),B(t),B(t)),\]
where $A=1+\phi(t)$ and $B=1-\phi(t)$. A straightforward calculation
of the $G^{00}$-component of the Einstein curvature tensor $G^{\mu\nu}$
yields\begin{equation}
G^{00}=-\frac{3}{4}\frac{\dot{B}(t)^{2}}{A(t)^{2}B(t)^{2}}.\label{eq:goocomp}\end{equation}
(The negative sign is due to a contraction of the curvature tensor
on the first and the forth index.) According to the Einstein field
equations, \begin{equation}
G^{\mu\nu}=8\pi T^{\mu\nu},\label{eq:efeq}\end{equation}
 $G^{00}$ is equivalent to an energy density $T^{00}$\begin{equation}
T^{00}=-\frac{3}{32\pi}\frac{\dot{B}(t)^{2}}{A(t)^{2}B(t)^{2}},\label{eq:tooincurv}\end{equation}
 which can be evaluated at our epoch $t=t_{0}$ as follows. We use
$\dot{\phi}(t)$ given in (\ref{eq:tdepscalepot}), the present value
of the Hubble constant $H(t_{0})=\frac{1}{1.3\times10^{28}{\rm cm}}$
in geometric units and an estimate of such a gravitational potential
of $\phi(t_{0})=-3.0\times10^{-5}$ as in \cite{ahluwgrcompl}, to
calculate $T^{00}$ in (\ref{eq:tooincurv}) to be:\begin{equation}
T^{00}=-0.06\frac{2.6\times10^{-66}}{{\rm cm}^{2}}.\label{eq:tooeval}\end{equation}
 This lies quite near at the Casimir energy density $\epsilon_{C}=-\frac{\pi^{2}\hbar c}{240\, d^{4}}$
\cite{casimirorig,deccaetalceexpannop} for a unit area and a distance
$d=1{\rm cm}$ written in geometric units, where Planck's constant
$\hbar$ becomes $\hbar^{gu}=2.6\times10^{-66}{\rm cm}^{2}$ and $c$
is put equal to one, \begin{equation}
\epsilon_{C}^{gu}=-0.04\frac{2.6\times10^{-66}}{{\rm cm}^{2}}.\label{eq:cegu}\end{equation}
 Although this calculation does not display the $d^{-4}$ dependence,
typical for the Casimir force, it points to a connection between a
gravitational environment described solely in terms of curvature by
the field theory of gravitation, i.e. general relativity, and a micro-physical
quantity containing Planck's constant. It should also be pointed out
that this calculation uses a part of the nonlinear Einstein field
equations, which is either often gauged away in calculations implying
flat background or is taken as instantaneous constraint rather than
representing a real field.

\section{Conclusion}

The result allows the speculation that Planck's constant might indeed
be connected to gravitational effects. It seems that our great teacher
Albert Einstein also was on the inside track in searching for field
theoretic explanations of quantum physical phenomena rather than spooky
ones - a fact not too often discussed in the celebrated Einstein year
2005.

\begin{acknowledgments}
J.L. gratefully acknowledges that his work has been supported by the
Hans Böckler Foundation.
\end{acknowledgments}

\end{document}